\documentclass{ifacconf}

\usepackage{graphicx}      % include this line if your document contains figures
\usepackage{natbib}        % required for bibliography

\usepackage{amsmath} % assumes amsmath package installed
\usepackage{amssymb}  % assumes amsmath package installed
\usepackage{siunitx}
\usepackage{booktabs}
\begin{document}
\begin{frontmatter}

\title{KIND: A Kalman-Inspired Adaptive Estimator for SRF Cavity Detuning\thanksref{footnoteinfo}} 

\thanks[footnoteinfo]{This work was supported by European Commission’s Horizon Europe Research and Innovation programme under Grant Agreement n°101131435. \copyright~2026 the authors. This work has been accepted to IFAC for publication under a Creative Commons Licence CC-BY-NC-ND.}

\author[First]{Andrei Maalberg} 
\author[First]{Axel Neumann} 
\author[First]{Pablo Echevarria}
\author[First]{Andriy Ushakov}
\author[First,Second]{Jens Knobloch}

\address[First]{Helmholtz-Zentrum Berlin, 14109 Berlin, Germany
  (e-mail: \{andrei.maalberg, axel.neumann, pablo.echevarria, andrey.ushakov, jens.knobloch\} @helmholtz-berlin.de).}
\address[Second]{Universität Siegen, 
   57068 Siegen, Germany (e-mail: jens.knobloch @uni-siegen.de)}

\begin{abstract}                % Abstract of 50--100 words
Superconducting radio frequency cavities with a high quality factor enable energy-efficient accelerator operation but are very sensitive to mechanical disturbances that detune their resonance. Accurate detuning estimation is therefore essential for efficient resonance control and stable beam conditions. This paper introduces Kalman-Inspired Neural Decomposition (KIND), a data-driven estimator that fuses a Dynamic Mode Decomposition model for stationary modal behavior with a Transformer-based predictor for transient dynamics. KIND further outputs learned uncertainty signals that indicate regime changes, enabling anomaly detection. Using operational cavity data, we compare KIND with a classical Kalman filtering baseline and discuss its potential as a foundation for future uncertainty-aware, forecast-based control.
\end{abstract}

\begin{keyword}
Data-driven modeling, state estimation, Koopman operator theory, uncertainty quantification, adaptive dynamical systems, Kalman filtering, SRF cavity detuning.
\end{keyword}

\end{frontmatter}
%===============================================================================

\section{Introduction}
The increasing push for energy-efficient operation of superconducting particle accelerators has led to the deployment of superconducting radio frequency (SRF) cavities with extremely high quality factors. While these high-Q systems reduce power consumption, they are also acutely sensitive to mechanical disturbances---such as microphonics or Lorentz-force detuning---that shift the resonance frequency of the cavity. Active detuning compensation therefore becomes essential to maintain stable beam conditions~\citep{Neumann_2010_cw_microphonics}.

Designing such compensation schemes begins with accurate detuning estimation. SRF cavities exhibit complex electromechanical behavior characterized by multiple narrow-band mechanical modes that couple to the electromagnetic field. Under stationary conditions, the detuning may be dominated by a few low-frequency modes (e.g., \SI{100}{\hertz} or below). However, transient events---such as liquid  helium fluctuations---can abruptly excite additional modes, fundamentally altering the system’s modal composition. These time-varying conditions violate the linear and stationary assumptions underpinning classical state estimators such as the Kalman filter.

The need for estimators that can adapt to evolving cavity dynamics has long been recognized within the accelerator community. \cite{Bellandi_2021_phd} emphasized that "the resonance controller should be designed to operate without complete a priori knowledge of the cavity mechanical dynamics and track eventual changes of the mechanical disturbances properties." Adaptive methods such as recursive least squares or extended Kalman filtering can partially tune model parameters online, but they remain structurally rigid---the number and coupling of mechanical modes must be fixed in advance. When these assumptions are violated, estimation quality deteriorates and divergence can occur, especially when the sign or magnitude of modal coupling is incorrect.

To address these limitations, we propose Kalman-Inspired Neural Decomposition (KIND)---a data-driven estimator that generalizes Kalman filtering principles to nonlinear and nonstationary SRF dynamics. KIND combines two complementary modules: a Dynamic Mode Decomposition (DMD)~\citep{Schmid_2010_DMD} model that captures the stationary modal structure, and a Transformer-based~\citep{Vaswani_2017_attention} predictor that models transient and nonstationary effects. Their outputs are adaptively fused through a learned uncertainty-dependent weighting analogous to the Kalman gain. Beyond producing the detuning estimate, each module also emits a latent consistency score, quantifying its confidence in the current input regime. These time-resolved uncertainty signals enable anomaly detection, highlighting moments when the observed dynamics deviate from previously learned behavior.

The present work focuses on the estimation and uncertainty modeling stage---a necessary precursor to future control schemes that exploit predictive uncertainty. Using operational SRF data exhibiting both stationary and transient regimes, we compare KIND against a classical Kalman filter as a representative stationary Gaussian baseline and analyze its self-assessed uncertainty behavior. The results show that KIND maintains estimation quality under modal transitions and that its uncertainty signals provide a reliable indicator of regime changes.

Our main contributions are as follows:

\begin{itemize}
    \item We introduce Kalman-Inspired Neural Decomposition (KIND), a hybrid DMD-Transformer estimator for nonlinear, nonstationary detuning dynamics.
    \item We compare KIND with a classical Kalman filter baseline and show that KIND preserves estimation quality under transient and distribution-shift conditions.
    \item We demonstrate that KIND’s learned uncertainty acts as intrinsic anomaly indicator, offering a bridge toward future uncertainty-aware control approaches.
\end{itemize}

\section{Related work}

Data-driven modeling has recently gained attention in the accelerator community as a means to improve the robustness of cavity diagnostics and detuning compensation.

\cite{Syed_2021_Koopman} introduced a Koopman-based Kalman filter for SRF systems, replacing a nonlinear cavity model with a learned linear embedding to enable real-time state estimation and fault detection. Their work demonstrated that latent-space filtering can capture nonlinearities while retaining Kalman interpretability. However, the linear latent model remains fixed and cannot represent transient shifts in modal composition.

\cite{Wang_2023_SRF} applied DMD to model cavity detuning and combined it with model predictive control. DMD effectively captures stationary modal dynamics but struggles under abrupt regime changes, where new modes become active or coupling coefficients vary.

In system identification and operator-theoretic modeling, combining linear spectral structures with neural components has proven effective for capturing dynamics beyond the reach of purely model-based approaches. Koopman-inspired neural models \citep{Wang_2023_KNF, Liu_2023_koopa} address distribution shifts by learning adaptive, data-driven operators on lifted state spaces.

KIND builds on these directions by fusing a DMD model of stationary behavior with a Transformer predictor for transients, blended through a learned uncertainty-dependent weighting analogous to the Kalman gain. Unlike prior methods, KIND also emits uncertainty signals that indicate when the input dynamics deviate from the learned ones, enabling anomaly detection. As such, KIND generalizes Kalman-style estimation toward adaptive, uncertainty-aware modeling of nonlinear dynamics.

\section{Background}

\subsection{Electromechanical model of an SRF cavity}

In this work, we consider the dynamics of an SRF cavity that arise from the coupling between its electromagnetic field and mechanical resonance modes. In this context, the electromagnetic response of the cavity can be represented as~\citep{Schilcher_1998_phd}

% --! rf cavity de -------------------------------------------------------------------
\begin{equation}
    %\begin{split}
        \frac{d}{dt}
        \begin{bmatrix}
            V_{T, I} \\
            V_{T, Q}
        \end{bmatrix}
        =
        \begin{bmatrix}
            -\omega_{1/2} & -\Delta\omega(t) \\
            \Delta\omega(t) & -\omega_{1/2}
        \end{bmatrix}
        \begin{bmatrix}
            V_{T, I} \\
            V_{T, Q}
        \end{bmatrix}
        %\\
        +\;
        \omega_{1/2}
        \begin{bmatrix}
            V_{F, I} \\
            V_{F, Q}
        \end{bmatrix},
    %\end{split}
    \label{eq:bkg_rf_cav}
\end{equation}

where $V_T$ and $V_F$ denote transmitted and forward cavity fields, $\omega_{1/2}$ is the half-bandwidth, and $\Delta\omega(t)$ represents the instantaneous detuning of the cavity resonance. Indices $I$ and $Q$ are the in-phase and quadrature components of the cavity field signals, respectively.

The detuning $\Delta\omega(t)$ is driven by the cavity’s mechanical modes, each modeled as a damped oscillator excited by the RF field gradient and, more generally, by external disturbances:

\vspace{-2mm}

\begin{equation}
    \frac{d^2}{dt^2} \Delta\omega_n
    +
    \frac{\omega_n}{Q_n} \frac{d}{dt} \Delta\omega_n
    +
    \omega_n^2 \Delta\omega_n
    =
    -k_n \omega_n^2 (V^2_{T, I} + V^2_{T, Q}),
    \label{eq:bkg_rf_mm}
\end{equation}

where $\omega_n$, $Q_n$ and $k_n$ are the frequency, quality factor, and coupling of the $n$-th mechanical mode. External excitations (e.g., microphonics) can be incorporated as additional forcing terms acting on the mechanical modes. The total detuning is the superposition

\vspace{-2mm}

\begin{equation*}
    \Delta\omega(t) = \sum_{n=1}^N \Delta\omega_n(t).
\end{equation*}

This interaction forms a nonlinear feedback loop: the RF drive, together with external disturbances, excites mechanical vibrations, which in turn shift the cavity’s resonant frequency.

\subsection{Kalman filtering as baseline estimation}

For small-signal or linearized approximations, detuning estimation is often formulated as a discrete-time linear state-space model

\vspace{-2mm}

\begin{equation*}
    x_t = A \, x_{t-1} + w_t, \quad z_t = Hx_t + v_t,
\end{equation*}

where $x_t$ is the latent cavity state, $z_t$ is the measured detuning, and $w_t \sim \mathcal{N}(0, Q)$, $v_t \sim \mathcal{N}(0, R)$ denote process and measurement noise, respectively. Following this, the Kalman Filter recursively updates state estimate $\hat{x}_t^-$ as

\vspace{-2mm}

\begin{equation}
    \hat{x}_t = \hat{x}_t^- + K_t \, (z_t - H \hat{x}_t^-).
    \label{eq:bkg_kalman_upd}
\end{equation}

The Kalman gain $K_t$ determines the confidence trade-off between the model prediction and measurement. The covariances $Q$ and $R$ are design parameters that encode epistemic and aleatoric uncertainty: large $Q$ values make the filter distrust its model (introducing epistemic uncertainty), while small $R$ emphasizes noisy measurements. In classical filtering, these uncertainties are homoscedastic, i.e., assumed constant over time. In contrast, complex systems such as SRF cavities exhibit operating regimes where uncertainty changes dynamically, e.g., when new mechanical modes are excited or coupling signs vary. KIND generalizes this notion of estimation by learning heteroscedastic uncertainty: it adapts its confidence weighting from data through learned, time-varying consistency scores. In this sense, it extends Kalman-style estimation toward adaptive, uncertainty-aware modeling of nonlinear system dynamics.

\section{Kalman-Inspired Neural Decomposition}

We now describe KIND architecture. KIND learns a data-driven dynamical model that maps a finite lookback window of past cavity detuning values to a forecast window of future evolution. 
Formally, for a discrete-time sequence $\{x_t\}_{t=1}^T$, the model operates on

\begin{equation*}
    X^{\text{back}} = [x_{t-T+1}, \ldots, x_t],
\end{equation*}

to predict

\begin{equation*}
    X^{\text{fore}} = [x_{t+1}, \ldots, x_{t+H}],
\end{equation*}

where $T$ and $H$ denote the lookback and forecast lengths, respectively. Following~\cite{Liu_2023_koopa}, we adopt $T = 2H$ to ensure that the lookback captures sufficient temporal context for stable forecasting.

\subsection{Lifting the dynamics}

Within each lookback window, we further segment the data into $\tau$-sized slices $\{X_j\}_{j=1}^{m}$, with $m = T / \tau$. These slices define the temporal resolution of the model: longer slices emphasize smoother, globally coherent structure, while shorter ones provide finer responsiveness to rapid transient dynamics.

\begin{figure}[!htb]
    \centering
    \includegraphics[width=0.75\linewidth]{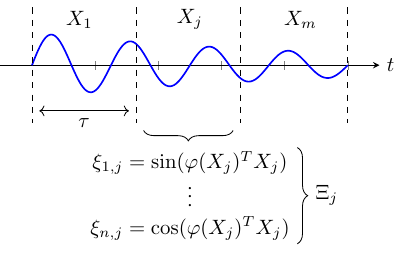}
    \caption{Lifting of time series data into latent embeddings. The lookback window is divided into $m$ $\tau$-sized slices, and each slice is evaluated through $n$ basis functions to form a lifted latent embedding $\Xi$.}
    \label{fig:method_timeseries_proc}
\end{figure}

Each slice $X_j \in \mathbb{R}^\tau$ is lifted into a latent representation using a set of basis functions $\mathcal{G} = \{g_i\}_{i=1}^n$ designed to approximate a Koopman-invariant subspace. 
The parameters of these basis functions are modulated by a kernel encoder $\varphi$, producing a latent variable matrix $V \in \mathbb{R}^{n \times m}$ via

\begin{equation*}
    v_{i,j} = \varphi(X_j)^\top X_j,
\end{equation*}

after which each basis function $g_i$ is evaluated to obtain latent embeddings $\xi_{i,j} = g_i(v_{i,j})$. Stacking these embeddings forms the lifted representation $\Xi \in \mathbb{R}^{n \times m}$, which evolves approximately linearly under a Koopman operator:

\begin{equation*}
    \mathcal{K}\mathcal{G}(V_j) \approx \mathcal{G}(V_{j+1}),
\end{equation*}

where $\mathcal{K} \in \mathbb{R}^{n \times n}$ advances the lifted state from slice $j$ to $j+1$.
Figure~\ref{fig:method_timeseries_proc} illustrates this lifting process.
 
\subsection{Stationary and transient Koopman operators}

A single Koopman operator $\mathcal{K}$ effectively models stationary, modal behavior but fails under transient or distributional shifts. 
To address this, KIND introduces two complementary branches:

\begin{itemize}
    \item a stationary branch, parameterized by a global Koopman operator $\mathcal{K}^{\text{stat}}$, representing stable modal structure; and
    \item a transient branch, with operator $\mathcal{K}^{\text{trans}}$, which adapts dynamically to local disturbances.
\end{itemize}

Each branch maintains its own latent embedding $\Xi^{b}$ ($b \in \{\text{stat}, \text{trans}\}$), ensuring independent representation and training. This decoupling allows the stationary dynamics to preserve long-term coherence, while the transient dynamics specialize in fast, short-lived perturbations. 
Practically, this modularity also supports architectural variations, e.g., allocating fewer basis functions to stationary dynamics and higher attention capacity to transients.

After evolution under their respective Koopman operators, the latent embeddings are decoded into the observable space using decoders $\psi^{b}$, producing:

\vspace{-2mm}
\begin{equation*}
    \hat{X}^{b}_{j+1} = \psi^{b}\!\left( \mathcal{K}^{b}\Xi^{b}_{j} \right),
\end{equation*}

and corresponding uncertainty estimates:

\vspace{-2mm}
\begin{equation*}
    \zeta^{b}_{j+1} = \zeta^{b}\!\left( \Xi^{b}_{j+1} - \mathcal{K}^{b}\Xi^{b}_{j} \right).
\end{equation*}

Intuitively, $\zeta^b$ decodes the prediction error in latent space—if the branch correctly models the observed input, its error and thus its uncertainty remain small. 
High uncertainty indicates that the branch does not "understand" the current input, signaling out-of-distribution or anomalous behavior. 
This mechanism makes $\zeta$ a powerful time-resolved diagnostic for model confidence that can eventually trigger an interrupt in case both branches fail.

\subsection{Uncertainty-driven blending}

KIND fuses the stationary and transient predictions using an uncertainty-weighted blending factor $\alpha_{j+1} \in [0,1]$, computed as

\begin{equation*}
    \alpha_{j+1} = \frac{\zeta^{\text{trans}}_{j+1}}{\zeta^{\text{trans}}_{j+1} + \zeta^{\text{stat}}_{j+1}}.
\end{equation*}

The fused prediction is then given by

\begin{equation*}
    \hat{X}_{j+1} = \alpha_{j+1} \, \hat{X}^{\text{stat}}_{j+1} + (1-\alpha_{j+1}) \, \hat{X}^{\text{trans}}_{j+1}.
\end{equation*}

This fusion is conceptually analogous to a Kalman update, where the blending factor $\alpha$ plays a role similar to the adaptive gain $K_t$. When the transient branch expresses high uncertainty (e.g., during stable operation), $\alpha$ increases, and the system relies on the stationary dynamics. Conversely, during a transient disturbance, the stationary branch becomes unreliable, $\alpha$ decreases, and the transient model dominates.

\begin{figure}[!htb]
    \centering
    \includegraphics[width=0.96\linewidth]{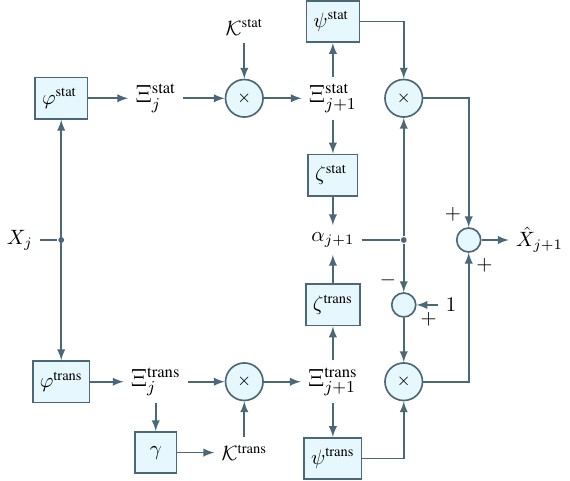}
    \caption{KIND architecture. For clarity, we depict operation on a single slice $j$. In practice, the model operates on all available slices simultaneously.}
    \label{fig:method_arch}
\end{figure}

The overall signal flow of the KIND architecture is summarized in Figure~\ref{fig:method_arch}. 
Each lookback slice $X_j$ is processed in two parallel branches: a stationary path with fixed Koopman operator $\mathcal{K}^{\text{stat}}$, and a transient path with a Transformer-inferred operator $\mathcal{K}^{\text{trans}}$. In the transient branch, the embeddings $\Xi^{\text{trans}}$ are processed by a network $\gamma$ that applies attention across the sequence of slices, enabling the construction of locally adaptive, time-varying operators that respond to short-term changes in dynamics. Both embeddings are decoded into predictions $\hat{X}^{b}_{j+1}$ and uncertainties $\zeta^{b}_{j+1}$, which are then fused via the uncertainty-driven gain $\alpha_{j+1}$. For brevity, the figure omits internal details of kernel encoders and basis functions, focusing instead on the primary computational blocks and signal flow.
 
\vspace{-0.7mm}

\subsection{Relation to Kalman filtering}

To emphasize this connection, the fused prediction can be written in Kalman-like form. Letting $\alpha = 1 - K_t$, and defining

\begin{equation*}
    \hat{x}^{\text{stat}} = \psi^{\text{stat}}(\mathcal{K}^{\text{stat}}\Xi^{\text{stat}}), \;\;
    \hat{x}^{\text{trans}} = \psi^{\text{trans}}(\mathcal{K}^{\text{trans}}\Xi^{\text{trans}}),
\end{equation*}

we obtain

\begin{equation*}
    \hat{x} = \hat{x}^{\text{stat}} + K_t \, (\hat{x}^{\text{trans}} - \hat{x}^{\text{stat}}),
\end{equation*}

which mirrors the classical Kalman update in (\ref{eq:bkg_kalman_upd}) with $\hat{x}^{\text{stat}}$ acting as the prior, $\hat{x}^{\text{trans}}$ as a learned "measurement," and $\zeta$ determining the adaptive gain. Unlike a conventional Kalman filter with fixed noise covariances $(Q, R)$---a homoscedastic assumption---KIND produces heteroscedastic, data-driven uncertainty that evolves with time and context. This enables both improved adaptation to regime changes and time-resolved anomaly detection capabilities.

\section{Experiments}

\subsection{Experimental setup}

The experiments use measured detuning data from an SRF cavity~\citep{Kamps_2025_gun}. The data was sampled at \SI{1}{\kilo\hertz} repetition rate. Relevant cavity parameters are summarized in Table~\ref{tab:exp_setup_cav_params}.

\begin{table}[!htb]
    \caption{SRF cavity parameters.}
    \label{tab:exp_setup_cav_params}
    \centering
    \begin{tabular}{lll}
        \toprule
        Frequency  & Loaded quality factor  &  Field gradient  \\
        \midrule
        \SI{1.3}{\giga\hertz}  & $4 \cdot 10^6$ &  \SI{9.5}{\mega\volt\per\meter} \\
        \bottomrule
    \end{tabular}
\end{table}

\begin{figure}[!htb]
    \centering
    \includegraphics[width=0.98\linewidth]{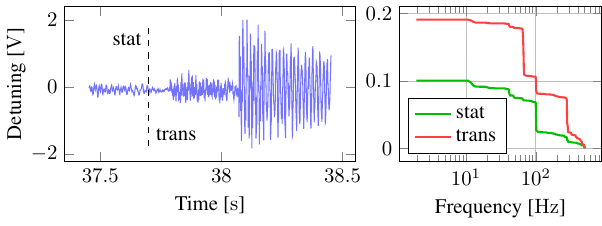}
    \caption{Measured SRF cavity detuning signal (left) and its integrated rms spectrum (right). Transient events excite additional mechanical modes that remain latent during the stationary regime.}
    \label{fig:exp_setup_meas}
\end{figure}

Figure~\ref{fig:exp_setup_meas} shows a representative detuning sequence together with its integrated rms spectrum. The time-domain trace includes both stationary intervals and transient activity. The spectral curve highlights pronounced steps associated with mechanical modes: a \SI{100}{\hertz} component dominates in stationary operation, while transient events excite additional modes (notably around \SI{64}{\hertz} and \SI{285}{\hertz}). This transient behavior prevents the use of the cavity with higher cavity fields, which negatively impacts electron beam quality.

\subsection{Kalman filter baseline}

To establish a physics-based reference, we designed a classical Kalman filter using the nominal detuning dynamics of the cavity. The mechanical subsystem was represented by three dominant modes at \SI{10}{\hertz}, \SI{40}{\hertz}, and \SI{100}{\hertz}, identified from the stationary portion of the measured data (see Figure~\ref{fig:exp_setup_meas}). The modes were modeled as discrete second-order state-space systems derived from (\ref{eq:bkg_rf_mm}). The corresponding mode frequencies $\omega_n$, quality factors $Q_n$ and coupling coefficients $k_n$ are summarized in Table~\ref{tab:exp_kalman_base_filter_param}. This model represents the expected detuning behavior during normal operation and forms the baseline against which both the transient data and KIND are evaluated.

\begin{table}[!htb]
    \caption{Kalman filter design parameters.}
    \label{tab:exp_kalman_base_filter_param}
    \centering
    \begin{tabular}{lllr}
        \toprule
        %& \multicolumn{3}{c}{Mode parameters}              \\
        %\cmidrule(l){2-4}
        Mechanical mode $n$        & $\omega_n$ & $Q_n$ & $k_n$  \\
        \midrule
        1                 & $2\pi\cdot 100$ & $1000$ & $1.0$ \\
        2                 & $2\pi\cdot 40$ & $400$ & $-1.0$ \\
        3                 & $2\pi\cdot 10$ & $100$ & $0.1$ \\
        \bottomrule
    \end{tabular}
\end{table}

The additional spectral components at \SI{64}{\hertz} and \SI{285}{\hertz} were intentionally excluded from this nominal model. They represent excitation modes that are not captured by the nominal low-order model and therefore cannot be accommodated without extending the underlying system identification procedure. This setup allows us to analyze how the Kalman filter behaves when confronted with unmodeled or misspecified dynamics---an essential point of comparison for data-driven approaches like KIND.

\begin{figure}[!htb]
    \centering
    \includegraphics[width=0.84\linewidth]{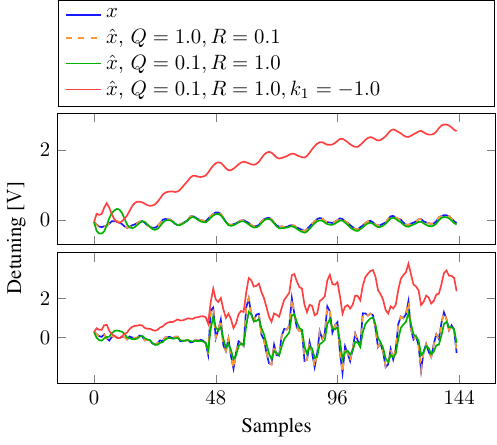}
    \caption{Kalman filter performance under different parameterizations in stationary (top) and transient (bottom) regimes. Model trust improves estimation only when structural assumptions hold; incorrect coupling signs produce estimation drift.}
    \label{fig:exp_kalman_base_performance}
\end{figure}

Figure~\ref{fig:exp_kalman_base_performance} illustrates the filter’s performance in both stationary (top) and transient (bottom) regimes under several covariance parameterizations $(Q, R)$ and one structural model error (incorrect coupling sign for the \SI{100}{\hertz} mode). With larger process noise $(Q=1.0, R=0.1)$, the filter relies more heavily on measurement updates, achieving stable tracking even under mild model mismatch. Reducing $Q$ while increasing $R$ $(Q=0.1, R=1.0)$ emphasizes model trust, leading to slower adaptation and, under coupling-sign inversion, to an evident estimation drift. This drift saturates at a steady offset, confirming that although the filter remains bounded, its internal model structure cannot self-correct.

Overall, these experiments show that tuning $Q$ and $R$ can temporarily mask modeling errors but not remove their cause. In contrast, KIND expresses uncertainty through a learned heteroscedastic blending mechanism, allowing the estimator to adapt its predictions under previously unseen dynamics without manual covariance retuning.

\subsection{KIND performance}

We next evaluate KIND on a detuning segment drawn from a transient regime dominated by the \SI{64}{\hertz} mechanical mode. In this example, the signal exhibits irregular, chopped oscillations throughout the lookback window, followed by a kick-like amplitude surge entirely contained within the forecast horizon. Such behavior requires not only capturing the modal frequency but also anticipating rapid amplitude changes.

As before, KIND operates in a lookback$\rightarrow$forecast manner: a 96-sample lookback window provides temporal context for a 48-sample open-loop forecast. This temporal context is important for identifying transient dynamical regimes and anticipating rapid amplitude changes that are not observable from instantaneous measurements alone. This contrasts with the recursive sample-by-sample updates performed by the Kalman filter under a fixed low-order state-space model.

\begin{figure}[!htb]
    \centering
    \includegraphics[width=0.88\linewidth]{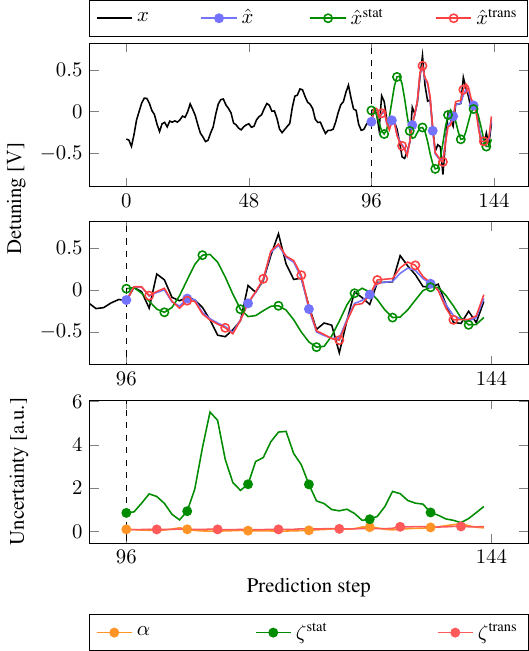}
    \caption{KIND forecast in a transient regime. The stationary operator diverges, while the transient operator captures the chopped \SI{64}{\hertz} oscillation and the kick-like amplitude rise. Uncertainty signals drive the blending toward the transient branch.}
    \label{fig:exp_kind_trans}
\end{figure}

Figure~\ref{fig:exp_kind_trans} shows the resulting forecast. The stationary operator continues the nominal low-amplitude dynamics consistent with the stationary modes (\SI{100}{\hertz}, \SI{40}{\hertz}, \SI{10}{\hertz}) and therefore diverges quickly from the ground truth. In contrast, the transient operator---whose dynamics are inferred by the Transformer network from the lookback window---closely reproduces both the oscillatory shape and the full amplitude growth of the kick.

\begin{figure}[!htb]
    \centering
    \includegraphics[width=0.85\linewidth]{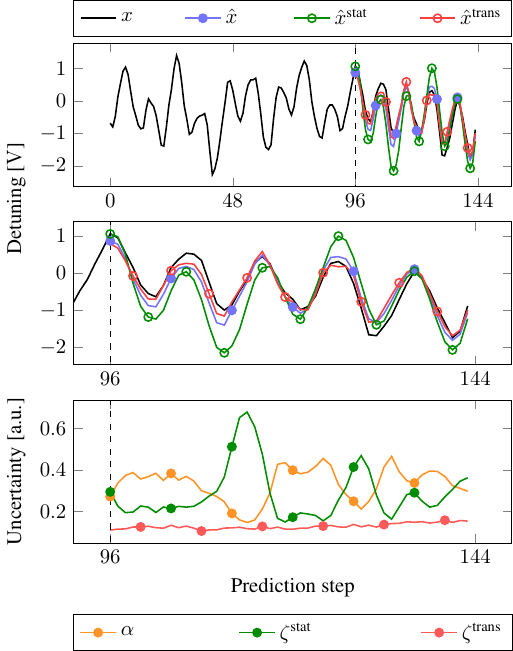}
    \caption{KIND forecast in a stationary-like \SI{100}{\hertz} regime. The stationary operator captures the frequency but overestimates the amplitude. The transient operator corrects this bias, and the uncertainties yield a balanced blend.}
    \label{fig:exp_kind_stat}
\end{figure}

Next, we evaluate KIND on a stationary-like segment of the measured detuning signal. Unlike the previous case, this region is dominated by a \SI{100}{\hertz} oscillation, with some contributions from other modes. This is precisely the scenario the stationary Koopman operator was trained for, although the plant model remains deliberately simplified (three-mode physics model). Figure~\ref{fig:exp_kind_stat} shows the ground-truth trajectory (144 samples) together with the KIND forecast on the 48-sample horizon. Two observations arise:

\begin{enumerate}
    \item The stationary operator correctly captures the dominant 100 Hz frequency but slightly overestimates its amplitude at several points. This reflects the mismatch between the simplified three-mode model and the true cavity dynamics. Importantly, the stationary uncertainty signal $\zeta^{\text{stat}}$ reacts appropriately: it rises at the time instants where the stationary forecast deviates from the measured amplitude.
    \item The transient (Transformer-based) operator produces a smoother, more accurate amplitude forecast, despite the absence of any strong transient in the input window. In this regime, its uncertainty 
$\zeta^{\text{trans}}$ remains low but non-zero, indicating that although the transient operator sees the regime as mostly familiar, it does not fully suppress its contribution.
\end{enumerate}

The resulting blending factor $\alpha$ therefore lies near 0.4, meaning that both branches contribute substantially. This is qualitatively different from the previous transient example, where 
$\alpha \approx 0$ and the transient operator dominated almost entirely.

Overall, this example demonstrates that KIND not only adapts under distributional shift, but also improves performance even in regimes where the stationary model is nominally applicable. The Transformer branch automatically compensates for deficiencies of the simplified physics model, while the uncertainty signals of both branches remain interpretable and well-correlated with forecast error.

\section{Discussion}

To ensure a controlled comparison with the Kalman filter, KIND's stationary branch was restricted to the same simplified three-mode model (\SI{100}{\hertz}, \SI{40}{\hertz}, \SI{10}{\hertz}). This highlights a common limitation of fixed low-order representations: they cannot capture the additional spectral content (\SI{64}{\hertz}, \SI{285}{\hertz},~…) present in measured SRF data. As a result, both the Kalman filter and the stationary operator exhibit similar degradation when the signal departs from nominal conditions. KIND compensates for these deficiencies through its transient branch, whose learned dynamics adapt to previously unseen amplitude and frequency patterns. Combined with uncertainty-based blending, this enables accurate forecasting on real data despite the underspecified stationary component. While incorporating a richer stationary basis could further improve performance and uncertainty calibration, this extension was intentionally left out of scope in order to preserve comparability with the Kalman baseline.

\section{Conclusion}

We presented KIND, a Kalman-inspired estimator that fuses physics-motivated Koopman operators with a data-driven transient correction mechanism. On measured SRF detuning data, KIND demonstrates improved robustness under transient and distribution-shift conditions and provides interpretable uncertainty signals that distinguish nominal and anomalous regimes. Future work includes automated discovery of dynamical regimes in SRF detuning signals and the integration of KIND forecasts into predictive control for resonance detuning compensation.

\section*{Declaration of AI Assistance}

During the preparation of this work the authors used OpenAI’s ChatGPT to assist with phrasing, structural editing, and refinement of exposition. After using this tool, the authors reviewed and edited the content as needed and take full responsibility for the content of the publication.

\bibliography{ifacconf}             % bib file to produce the bibliography

\end{document}